\documentclass{aa}  
\usepackage{graphicx}
\usepackage{txfonts}
\begin{document}
\title{The impact of the dark matter-gas interaction on the collapse behaviour of spherical symmetric systems}


\author{J. S. Klar
  \and
  J. P. M\"ucket
}


\institute{
  Astrophysikalisches Institut Potsdam\\
  \email{jklar@aip.de, jpmuecket@aip.de}
}

\date{Received 26 February 2008; Accepted 4 April 2008}

\titlerunning{The impact of the DM-gas interaction on the collapse behaviour
  of spherical symmetric systems}


\abstract
{If the gas in the evolving cosmic halos is dissipating energy (cooling) then due to the variation of the gravitational potential the dark matter halo also undergoes a compactification. This is well-known as Adiabatic contraction (AC).} 
{Complementary to the AC we investigate the resulting dynamical behaviour of the whole system if the backreaction of the AC of DM onto the gas is taken into account.}
{In order to achieve sufficient high resolution also within the central halo region, we use a crude fluid approximation for the DM obeying the adiabatic contraction behaviour. Further, we restrict ourself to spherical symmetry and vanishing angular momentum of the studied matter configurations. The computations are done using a first-order Godunov type scheme.} 
{Our results show that the dynamical interaction between gas and DM may lead to significant shorter collapse times.  If the gas cools the dynamical behaviour of the whole system depends strongly on the shape of the initial density profile. }
{Our findings indicate that for a certain mass range of halo configurations the dynamical interaction between gas and DM might be important for the halo evolution and must be taken into account.}


\maketitle

\section{Introduction}

During the past decade considerable progress has been achieved in understanding the structure formation process in the universe. In particular, the dynamics and evolution of the dissipationless component, i.e. the formation and evolution of the dark matter (DM) halos was studied in great detail.
The outcome of all the cosmological simulations is that the dark-matter
component forms a nearly stable halo with an almost universal density shape
which is well approximated by the profile fit given by \cite{navarro97} (NFW
profile). 

However, the density profiles derived from the observed rotation curves of galaxies (in particular, from that of LSB galaxies) 
indicate on much shallower profiles than the NFW profile or indicate even on a core structure (\cite{marchesini02}, \cite{gentile04}, \cite{kuzio06}, \cite{salucci07}). 
Therefore, the question arises whether the DM halo profiles have really an
universal shape and in particular, which mechanisms or processes could then
influence or change the DM density profile towards the observed ones. The structure of 
the inner DM density profile (cuspy or core like) is important with respect to various processes 
within the innermost parts of galaxies and clusters of galaxies. 
The galaxy-galaxy lensing is a very promising observational method to map the dark matter distribution in galaxy halos. For a successful reconstruction the knowledge of the density profile for a single DM halo is necessary, however (\cite{kleinheinrich06}). 
The innermost DM density profile, i.e. how much concentrated the DM is within the central parts, is of particular interest if
considering the consequences of possible dark matter annihilation and estimating the $\gamma$-ray luminosity of Galactic halos (see, e.g., \cite{stoehr03}, \cite{acasibar06}, \cite{diemand07}). 

Contrary to the DM, the baryonic fraction, ($M_b \approx \, 0.2 
\, M_{DM}$)
is able to dissipate energy, may cool, and eventually forms a configuration
much more concentrated than the surrounding DM. Due to the gravitational
coupling of both components the DM responds dynamically on the
compactification of the baryonic gas and the density shape of the DM is
changing. That effect was first studied by \cite{eggen62} using the model of
adiabatic contraction (AC). \cite{zeldovich80} used the approach of adiabatic
invariants to consider an analogous situation with respect to mass constraints on Leptons. A well elaborated form of the
AC model was given by \cite{blumenthal86}. The model was later investigated and tested by \cite{ryden87}, \cite{cardone05}, \cite{sellwood05}, \cite{vasiliev06}, \cite{dutton07} and has been
partly modified by  \cite{jesseit02}. In particular, it was considered within a cosmological
context by \cite{gnedin04}.
From the above studies, it is not clear however which processes are mainly
influencing the final DM profile.  \cite{gnedin04} have investigated how
mergers and different cooling periods for the baryonic gas act on the DM
dynamics. However, within mergers mainly the interaction of the DM particles
is important. In the very inner region the steepest matter distribution
defines the DM profile. It has been mentioned that the formation of the final
DM profile might depend on the relation of cooling and dynamical times. The AC
model assumes a relatively slow (adiabatic) change of the potential. Since the
convergence of the simulations within $r<< r_{Virial}$ is still not proven, it
is difficult to make a final conclusion about the density evolution in the
central region. However, this is the region which is most affected by any
concentration of the gas in the result of cooling processes. 

We want to consider the relatively artificial situation of a spherical
symmetric matter distribution with fixed masses and limited spatial
extension. The mass relation of gas to DM is fixed as 0.25. 
This ratio on average varies for the observed cosmic structures (galaxies, clusters, dwarf galaxies, etc.) very much since it depends on the particular conditions of the their formation. It is also correlated with the total mass of objects. Thus, the assumed unique mass ratio of about 0.25 is taken in our considerations as an arbitrary value.
In order to
explore the dynamical interaction between the DM and gas we start from a
stable stationary configuration and then we let the gas cool. Thus, no effects
of particular  matter infall or merger-like mass growth is considered. To
enhance the resolution considerably we use for the description of the DM a
crude gas/fluid approximation assuming that the violent relaxation processes
are fast enough to satisfy a nearly Maxwellian velocity distribution of the DM
particles at each time and within each shell, provided  a sufficiently coarse
grained shell distribution is considered. A similar treatment of
  the DM was used before in \cite{chieze} and \cite{teyssier}. Therefore, we describe at high
accuracy the interaction of two gases/fluids which interact only
gravitationally, and only the gas component is able to dissipate/cool. This
allows to consider not only the dynamical response of the DM on the baryonic
cooling as described by the AC models but vice versa the back-reaction on the
gas dynamics, too. 

Recently, \cite{conroy07} have investigated which kind of energy input processes in a system comparable to clusters of galaxies may prevent the cooling catastrophe. Their outcome is that the common action of various heating processes may be able to stabilize the system over cosmic time. However, a fine tuning is necessary and the whole system is very sensitive to the parameters. They have considered the dynamics of the gas assuming the DM potential to be static. Our studies demonstrate that the regard of the gravitational interaction between gas and DM increases the tendency to an onset of the cooling catastrophe. 
The paper is organized as follows: In the next section we introduce our basic
assumptions and give the main equations for the considered system. The
numerical treatment is given in section 3. Section 4 contains the results for
our various considered cases where the time evolution of the DM halo+gas
system is followed if the gas cools. A discussion and summary of these results
are given in section 5. For better systematics, the details of the used
numerical algorithm are described in the appendix. 

\section{Basic equations}
This section reviews our main assumptions and the basic equations describing
the system of a dark matter halo with embedded gas.

\subsection{Dynamics of the baryonic gas}

The baryonic fraction of the halo is described as an ideal gas. The
hydrodynamical equations for the density $\rho_G$, 
the radial momentum density $\rho_G \, u_G$ and the 
energy density $\rho_G \, \varepsilon_G = \rho_G\, u_G^2 / 2 + \varepsilon_{G,T}$ of 
a gravitationally bound gas sphere (we assume spherical symmetry) are:
\begin{eqnarray}\label{gas}
\frac{\partial}{\partial t} \, \rho_G
+ \frac{1}{r^2} \frac{\partial}{\partial r} \left[ r^2 \rho_G \, u_G \right] &=& 0
\nonumber \\
\frac{\partial}{\partial t} \, \left( \rho_G \, u_G \right) 
+ \frac{1}{r^2} \frac{\partial}{\partial r}\left[ r^2 \, \rho_G \, u_G^2 \right]
+ \frac{\partial p_G}{\partial r}
&=& - \, \rho_G \, \frac{\partial \Phi}{\partial r}
\nonumber \\
\frac{\partial}{\partial t} \, \left(\rho_G \, \varepsilon_G\right)
+ \frac{1}{r^2} \frac{\partial}{\partial r} \left[ r^2 u_G \left( \rho_G \, \varepsilon_G + p_G \right)\right]
&=& - \, \rho_G \, u_G \, \frac{\partial \Phi}{\partial r} 
- \Lambda
\end{eqnarray}
with the gravitational potential $\Phi$ given by the Poisson equation 
and the cooling function $\Lambda$. Pressure $p$, internal energy $\varepsilon_G$ 
and density $\rho_G$ are related to each other by the polytropic equation of state:
\begin{equation}
\varepsilon_{G,T} = \frac{p_G}{(\gamma - 1) \, \rho_G}
\end{equation}

The gravitational potential is determined by the density distributions of the gas $\rho_G$ and of the DM $\rho_{DM}$ via the 
Poisson equation:
\begin{equation}\label{g}
\frac{1}{r^2} \frac{\partial}{\partial r} 
\big[ r^2 \, \frac{\partial \Phi}{\partial r} \big] = 4 \pi \, G \,
\big[ \rho_G + \rho_{DM} \big] \;.
\end{equation}
 
To consider typical energy dissipation in the gas we assume cooling by recombination as given in \cite{black}. We further assume that the gas consists of hydrogen, only. In that case we have
\begin{eqnarray}\label{black}
\Lambda &=& 2,85 \cdot 10^{-27} \, T^{1/2} \nonumber\\
&& \cdot \left( 5,914 - 0.5 \ln T + 0,01184 \, T^{1/3}\right) \, n_e \, n_{H^+}
\end{eqnarray}
where $n_e$, $n_{H^+}$ denote the number densities of electrons and protons and $T$ being the temperature. 
We consider only collisional ionization and suppose ionization equilibrium
\begin{eqnarray}\label{iogg}
\Gamma_{EH}(T) \, n_H \, n_e = \alpha_H(T) \, n_{H^+} \, n_e
\end{eqnarray}
where $\Gamma_{EH}(T)$ and $\alpha_H(T)$ is the corresponding ionization and recombination rate, respectively (\cite{black}).

Since the gas is assumed to consist of hydrogen only and and due to electric charge conservation 
the relation between the number densities $n_e$, $n_{H^+}$ and $n=\rho_G/m_H$ can be determined straightforward as:
\begin{eqnarray}\label{cool3}
n_e = n_{H^+}= n \, \frac{\Gamma_{EH}}{\Gamma_{EH} + \alpha_H} 
\end{eqnarray}
and (\ref{black}) can be written in terms of $n$ and $T$ only.

In order to simplify the calculations, we restrict ourselves to the above 
  cooling mechanism troughout. One should be aware however that within large
  mass systems like galaxies and clusters of galaxies the dominant cooling
  processes are helium and heavy element line cooling and Bremsstrahlung. In a
  realistic treatment, the cooling function is a composite of the various
  contributions the importance of which depends on the precise temperature and density conditions 
at each moment for the considered volume.

\subsection{The dark matter dynamics - a fluid approach}

The dynamics of the dark matter is given by the collissionless
Boltzmann-equation for the one-particle phase-space density
$f({\mathbf x},{\mathbf v},t)$
\begin{equation}
\frac{\partial f}{\partial t}  
+ \sum_i v_i \, \frac{\partial f}{\partial x_i} 
+ \sum_i \frac{\partial \Phi}{\partial x_i} \, \frac{\partial f}{\partial v_i} 
= 0
\end{equation}
along with the Poisson equation for the mean gravitational potential $\Phi$
(see Eq. \ref{g})

To obtain fluid-like equations of motion for the dark
matter we compute the moments with respect to the velocities $v_i$ up to the
second order using the CBE as described in \cite{bt} and take into account
that the velocities vanish at $r\to\infty$. Our main approximation is to
assume that the 3rd order moments vanish. In particular, this is true if the
velocity distribution is locally Maxwellian. Furthermore, we assume the
  velocity dispersion to be isotropic. Under these crude assumptions one
gets the system of equations to be closed. In addition to the Jeans equation
we get then an energy equation.

If we assign formally the quantity 
\begin{equation}
p_{DM} = \frac{1}{3} \; \rho_{DM} \, \sigma^2_{DM} = \; \rho_{DM} \, \sigma^2_{DM,r}
\end{equation}
to a pressure then we have a full analogy to the gas equations. Making such an assumption we seem  to suppose a kind of collision between the DM particles. However, if considering sufficiently coarse grained shell distributions for halos in n-body simulations then one observes nearly Maxwellian velocity distributions within each mass shell (see, e.g. \cite{hoeft04}). This is most probably the result of sufficiently fast violent relaxation during the collapse of the considered systems. With other words, the mechanism of violent relaxation (\cite{lb}, \cite{kull}) has to
redistribute the velocities of the dark matter faster then the dynamical
timescale. For the initial configuration, which we will start from, this process of relaxation leading to some equilibrium and to corresponding velocity distribution is assumed to be completed at the moment $t=0$. Further on, the time-dependent evolution happens slowly enough that shell crossing is negligible. 

We want to use that approximation to study the interplay between the gas and DM dynamics if energy dissipation in the gas happens. After transformation to spherical coordinates and restricting ourself to isotropic velocity dispersion (i.e., $\beta=0$) we obtain the following system of equations:
\begin{eqnarray}\label{dm1}
\frac{\partial}{\partial t} \, \rho_{DM}
&+& \frac{1}{r^2} \frac{\partial}{\partial r} \left[ r^2 \rho_{DM} \, u_{DM}
\right] 
\;=\; 0 
\nonumber \\
\frac{\partial}{\partial t} \, \left( \rho_{DM} \, u_{DM} \right) 
&+& \frac{1}{r^2} \frac{\partial}{\partial r} 
\left[ r^2 \,  \rho_{DM} \, u_{DM}^2  \right] 
+ \frac{\partial}{\partial r} \frac{ \rho_{DM} \,\sigma^2_{DM}}{3}
\nonumber \\
&&\qquad\qquad\qquad\qquad
\;=\; - \;\rho_{DM} \, \frac{\partial \Phi}{\partial r} 
\nonumber \\
\frac{\partial}{\partial t} \, \left(\rho_{DM} \, \varepsilon_{DM}\right)
&+& \frac{1}{r^2} \frac{\partial}{\partial r} \left[ r^2 u_{DM} \left( \rho_{DM} \,
    \varepsilon_{DM} + \frac{ \rho_{DM} \,\sigma^2_{DM}}{3} \right)\right] 
\nonumber \\
&&\qquad\qquad\qquad\qquad
\;=\; - \; \rho_{DM} \, u_{DM} \, \frac{\partial \Phi}{\partial r} \; ,
\end{eqnarray}
with the total energy density
$\rho_{DM} \,\varepsilon_{DM} =  \rho_{DM} \, v_r^2 / 2
+ 3 \, \rho_{DM} \, \sigma^2_r / 2$.

Equations (\ref{dm1}) together with the Poisson equation provide a closed system for
the dynamics of the dark matter. As mentioned above these equations exhibit a strong analogy to the hydrodynamic equations as described before.
Using the continuity equation we can perform the energy equation in Eq.(\ref{dm1}) to
\begin{equation}\label{dm4}
\frac{D}{Dt} \log\left({\frac{\rho_{DM}}{\sigma^3_{DM}}}\right) = \left[\frac{\partial}{\partial t} + u_{DM} \frac{\partial}{\partial r}\right] \log\left({\frac{\rho_{DM}}{\sigma^3_{DM}}}\right) = 0
\end{equation}
The quantity $\tau = \rho_{DM}/\sigma_{DM}^3$ represents the phase space
density (see \cite{taylor01}) and the logarithm of $\tau$ is a measure for the
entropy. Eq. (\ref{dm4}) states that the entropy is constant along the flux
lines, and therefore also $\tau=const.$. Thus, the assumptions on the
velocity distribution of the DM yields an effective equation of state (e.o.s.) which describes the
process of adiabatic contraction of the DM sufficiently well. For the pressure
analog $p_{DM} = \rho_{DM}\sigma_{DM}^2$ we get then the relation $p_{DM}
\propto \rho_{DM}^{5/3}$ formally identically to the adiabatic e.o.s. for an
ideal gas. 

Note, for the static case the Eq. (\ref{dm4}) is identically fulfilled, and neither any restrictions on the e.o.s. nor any explicite expression for the e.o.s. can be obtained.

\section{Numerical calculations}

\subsection{The initial configuration of the DM-gas system}
For the initial distribution we chose a 
polytropic gas sphere (see \cite{bt}) of finite extension. We combine this with
\begin{enumerate}
\item a spherical DM distribution obeying an effective e.o.s. with $\gamma_{DM}=5/3$
\item a dark-matter distribution following a  NFW-profile
\end{enumerate}

According to the previous findings, for the dark matter and for the gas may be assumed
\begin{equation}
p_{DM} = \kappa_{DM} \cdot \rho_{DM}^{\frac{5}{3}}
\quad \mbox{and} \quad
p_{G} = \kappa_G \cdot \rho_G^{\gamma} \:.
\end{equation} 
with the constants $\kappa_{DM}$ and $\kappa_G$.
The initial profiles itself are computed for hydrostatic equilibrium and it's analog for the DM
\begin{equation}\label{equi}
\frac{d p_{DM}}{d r} = 
\, - \rho_{DM} \, \frac{d \Phi}{d r} 
\quad \mbox{respectively} \quad
\frac{d p_G}{d r}  = - \,\rho_G \, \frac{d \Phi}{d r} 
\; .
\end{equation}
Together with the gravitational acceleration 
$g = - d \Phi / d r$ computed by Equation (\ref{g}), 
one gets the following system of equations
\begin{eqnarray}
\frac{d}{d r} \, \rho_{DM}
&=& \frac{g}{\frac{5}{3} \, \kappa_{DM}}\, \rho_{DM}^{2 - \frac{5}{3}}
\; ; \quad
\frac{d}{d r} \, \rho_G 
\;=\; \frac{g}{\gamma \, \kappa_G}\,  \rho_G^{2 - \gamma}
\nonumber\\
\frac{d}{d r} \left( r^2 g \right) 
&=& - \,4 \,\pi \, G \, \left( \rho_{DM} + \rho_G \right)\, r^2 
\end{eqnarray}
which can be easily solved by a Runge-Kutta algorithm. The
central density of the gas and of the dark matter, the constants 
$\kappa_{DM}$ and $\kappa_G$ introduced above, and the ratio of specific heats for the gas $\gamma$ are free parameters. The central density gradients are set equal to zero. If we want the configuration of both, the dark matter and the gas, to be bound at the same finite radius, then we have to satisfy the condition 
\begin{equation}\label{chi}
\kappa_{DM} = \kappa_G \, \left[ \rho_{G}^{\gamma_G - \frac{5}{3}}\right]_{(r=0)}
\; ,
\end{equation}
This eliminates one free parameter. For a better interpretation 
we chose the halo mass $M_H$, 
the cut-off radius of the profile $R_H$ and the ratio of specific heats
 $\gamma$ instead of the above introduced free parameters .
As an additional constrain we fix the ratio of total dark matter mass 
and total gas mass to be 0.25.

As an alternative we consider the initial DM density distribution realizing a
NFW density profile. In that case, the density profile for the baryons is computed as above, but taking into account the given NFW-profile for 
the DM within the Poisson equation:
\begin{equation}
\rho_{NFW} = \frac{\delta}{r \, \left( r + r_s \right)^2}
\end{equation}
Here, the normalization for $\delta$ is chosen in a way that the ratio of the total dark 
matter mass and the total gas mass is fixed as 0.25. The characteristic radius is
set as an additional parameter. Thus, the equations for the 
the initial gas profile are:
\begin{eqnarray}
\frac{d}{d r} \, \rho_G 
&=& \frac{g}{\gamma \, \kappa_G}\,  \rho_G^{2 - \gamma}
\nonumber\\
\frac{d}{d r} \left( r^2 g \right) 
&=& - \,4 \,\pi \, G \, \left( \rho_{NFW} + \rho_G \right)\, r^2 
\end{eqnarray}
In a second step $p_{DM}$ is determined to keep the system initially in
hydrostatic equilibrium (Eq.(\ref{equi})).

\subsection{The dynamical evolution}

Once the system is forced to leave the hydrostatic equilibrium it's further evolution is determined by the full timedependent hydrodynamic and Poisson equations. In our case, the subsequent cooling is responsible for this process.
In order to compute the evolution of the system, we choose a Lagrangian scheme which is 
equivalent to the above Eulerian scheme. This has the benefit of conserving
the mass automatically, and the treatment of the outer halo boundary is much
simpler. We use the above discussed formal analogy to define a pressure for the DM
as $p_{DM} = \rho_{DM} \, \sigma^2_{DM} \, / 3$. This allows us for using formally the same equations for the dark matter as for the gas.
\begin{eqnarray}
\frac{\partial \tau}{\partial t} 
&=& \frac{\partial}{\partial m} \left( r^2 \,u\right)
\; ;\quad
\frac{\partial u}{\partial t}
\;= \;
- r^2 \frac{\partial p}{\partial m} + g 
\; ;\nonumber\\
\frac{\partial \varepsilon}{\partial t}
&=& 
- \frac{\partial}{\partial m} \left( r^2\, u\, p \right) 
+ u\, g + \tau \, \Lambda 
\; ;\quad
\frac{\partial r}{\partial t} 
\;=\; u
\; ;\nonumber\\
\tau &=& \rho^{-1}
\; ;\quad
\varepsilon \;=\; \frac{u^2}{2} + \frac{p \, \tau}{\gamma - 1}
\; ;\quad
\, m(r) = \int_{0}^r r^{\prime 2} \,\tau^{-1} \, \mbox{d}r^{\prime}.
\end{eqnarray}
The gravitation is the only coupling between the two kinds of halo matter,
all other calculations may be performed independently for each component using 
identical routines. The intercell fluxes are computed using a 
first-order Godunov type scheme. We modified the original (second-order) 
MUSCL scheme presented in \cite{muscl}. In our case, due to the achieved high resolution the first-order scheme is sufficient, and encountering strong shocks is not expected.
The Algorithm is in detail described in Appendix \ref{algo}.

The sources consisting of gravitational acceleration and cooling
are calculated in each cell center. The gravitational acceleration at a given 
radius, is computed using the sum of the enclosed dark matter and gas mass:
\begin{equation}\label{grav}
g = - \frac{4 \, \pi \, G\, \left( m_{DM} + m_G \right)}{r^2}
\end{equation}
The cooling function $\Lambda \, \tau $ is calculated according to
Equations (\ref{black}) and (\ref{cool3}).

Instead of the physical quantities we use a system of dimensionless quantities
based on the initial central gas density $\rho_0$ and the velocity 
$u_0 = c_0 / \sqrt{\gamma_G}$ , with $c_0$ being the initial central 
speed of sound of the gas. The dimensionless quantities are computed as:
$ \bar{\rho} =  \rho / \rho_0, \; \bar{u} = u / u_0, ...$
Radius and time are tranformed according to 
\begin{equation}
r_0 = \sqrt{u_0^2 / (4 \pi G \rho_0)}\; , \quad
t_0 = r_0 / u_0 =\sqrt{1 / (4 \pi G \rho_0)}
\end{equation}
The dynamical equations remain unchanged under this transformation, and 
Eq.(\ref{grav}) simplifies to:
\begin{equation}
\hat{g} = - \frac{\left( \hat{m}_{DM} + \hat{m}_G \right)}{\hat{r}^2}
\; .
\end{equation}
Note, without cooling the system is free scalable and the actual dimensions of the
halo have no impact on the dynamics. The cooling function has to 
be computed 
using the physical density and temperature and must then be transformed according
to
\begin{equation}\label{realcool}
\hat{\Lambda} \, \hat{\tau} = \frac{\tau_0 \, r_0}{u_0^2} \, \Lambda \, \tau
\; .
\end{equation}

\section{Results}
\begin{figure}
   \centering
   \includegraphics[width=\columnwidth]{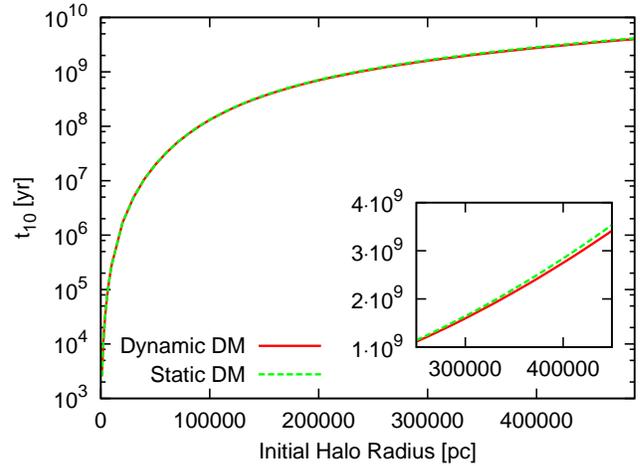}
      \caption{Time $t_{10}$ when the central gas density reaches 10 times the initial central density as function of the initial boundary radii at mass $M_H = 10^{11} M_{\odot}$. Note, for fixed mass the $R_H$ is a measure for the mean density.}
      \label{time-11}
   \end{figure}

We performed a series of computations starting with initial stationary
configurations as described above.  First, it was proven that these configurations (if allowing for their time evolution) are stable against small perturbations as long no cooling takes place.
Then we allowed the gas of the system to
cool. This drives the system out of the initial hydrostatic equilibrium. The
further thermal and dynamical evolution is governed by the system of Equations
(\ref{gas}), (\ref{g}) and (\ref{dm1}). For large
halo masses comparable with those of galaxies and more massive objects, the
system always collapses within a period comparable with the free fall time of
a sphere with mean density $\rho_{average}\approx M_H/R_H^3$. The
Fig. \ref{time-11} shows a measure of the collapse time in dependence on the
initial boundary radii at fixed gas + DM halo mass, i.e. as function of the mean density. 

If the contraction of the DM fraction is taken into account, the collapse time is marginally smaller. We want to get information about the action of the contracting DM onto the stability behaviour of the whole system. Therefore, we consider the behaviour of halos with masses of about $M_6=10^6 M_{\odot}$. Those masses are stable against a short cooling period leading to a finite contraction whereas subsequent cooling leads also to an unlimited collapse. However, as can be seen below, in that case a period of quasi-stability or delayed contraction may occur. This behaviour is more sensible with respect to the dynamics of the DM.
In that sense halos with $\approx M_6$ represent a limiting case and the action of the co-collapsing DM can be studied in more distinctness. 

In order to study the influence of the initial density profile we have run the
time-dependent computations for a set of different radii of the initial gas
configuration. The NFW profile is formally extending till infinity but due to
the $r^{-3}$ dependence at the outer region it decreases very fast. We have
located the border of the initial gas sphere well within the region where
$\rho_{DM} \to 0$ but varied it extension within a relatively narrow range
($R_{H,G}=[20,28,34] pc$). For the case of an initially "polytropic" DM
distribution we can achieve identical outer boundaries for the DM and gas. In
this case we have varied the initial radii as $R_{H,G}=[16,20,24]pc$. 
The particular values for $R_{H,G}$ were chosen to demonstrate the possible qualitatively different results.

\begin{figure}
   \centering
   \includegraphics[width=\columnwidth]{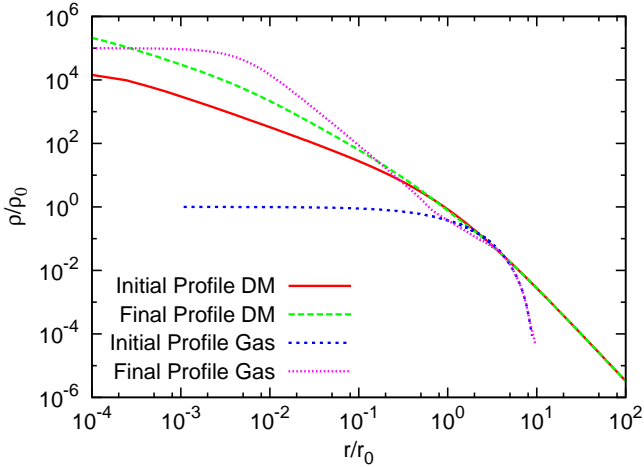}
      \caption{Initial and final density profiles for DM and gas. The final
   density profiles are given at the moment when the central gas density
   reaches the value $10^5$. The initial DM profile follows a NFW profile.
   ($R_H = 20 pc$)
   }
      \label{profile-nfw}

   \end{figure}

\begin{figure}
   \centering
   \includegraphics[width=\columnwidth]{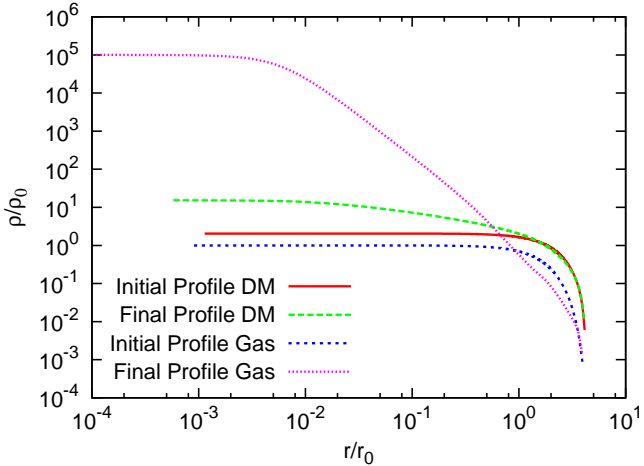}
      \caption{ Same as in Fig. \ref{profile-nfw} but for an initial
   "polytropic" DM distribution. ($R_H = 16 pc$)}
      \label{profile-poly}
   \end{figure}

Figures \ref{profile-nfw} and \ref{profile-poly} show the typical evolution of
the density profiles for gas and DM for an initial NFW profile and a
"polytropic" profile (with formally $\gamma_{DM}=5/3$) for the
DM distribution. When the central gas density increases by 5 orders of
magnitude the central DM density is increasing by one order of magnitude in
both cases. The density profiles are steepening with time.


It is obvious that the larger the initial radii are the smaller the initial
central density is. The time evolution of radii containing a fixed amount of mass ($r(0.5 M), r(0.1 M), r(0.01 M)$) is given in the Figs. \ref{mass-nfw} and \ref{mass-poly} for NFW and "polytropic" initial DM distributions respectively.

   \begin{figure*}
   \centering
   \includegraphics[width=16cm]{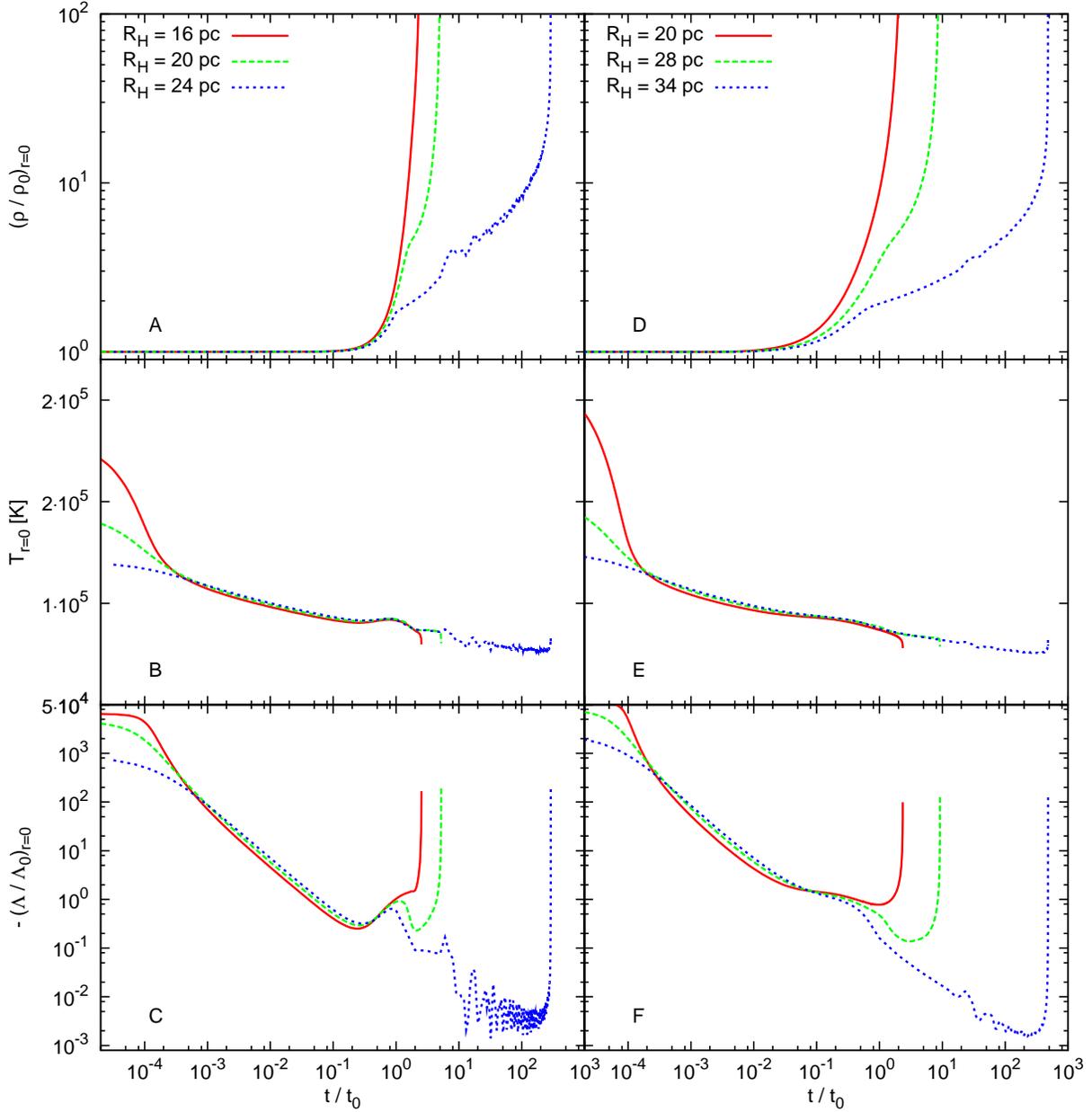}
      \caption{Time evolution of central gas density (A,D), central temperature (B,E) and of the absolute value of the cooling function (C,F). The left column shows the quantities for the case of an initial "polytropic" DM distribution, the right column shows the behaviour for an initial NFW profile of DM distribution.}
      \label{sextett}
   \end{figure*}

 \begin{figure}
   \centering
   \includegraphics[width=\columnwidth]{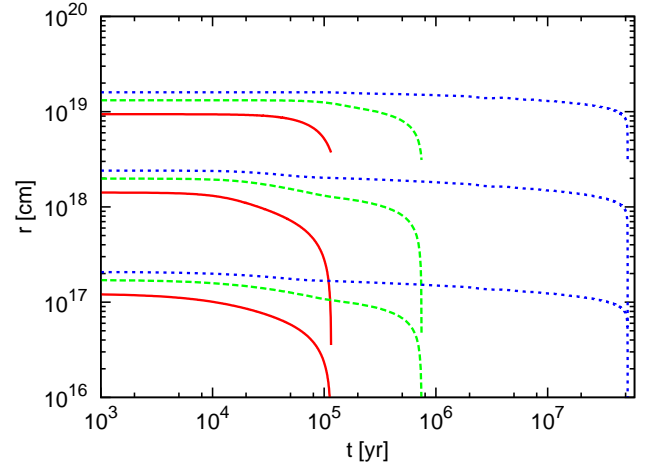}
      \caption{The time evolution of different radii containing fixed a fraction of mass (top to bottom $r(0.5 M), r(0.1 M), r(0.01 M)$) for various initial halo boundaries labeled by line style as in Fig. \ref{sextett} for an initial NFW profile of the DM.}
      \label{mass-nfw}
   \end{figure}
   \begin{figure}
   \centering
   \includegraphics[width=\columnwidth]{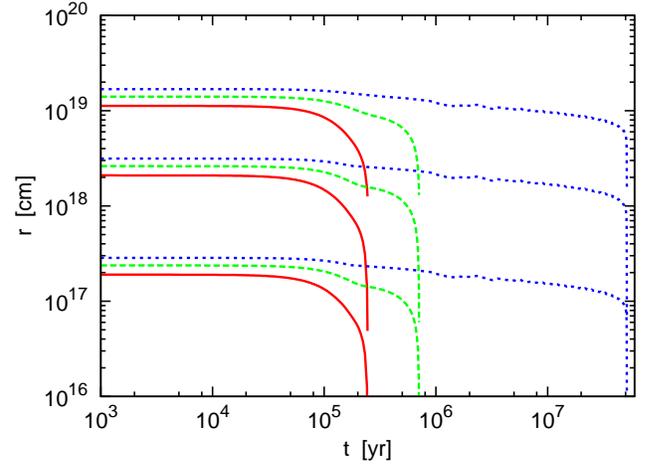}
      \caption{The same as in Fig. \ref{mass-nfw} but for an initial "polytropic" DM distribution.}
      \label{mass-poly}
   \end{figure}


The time evolution of the central densities shows the most pronounced differences with respect to the chosen initial parameters. Fig. \ref{sextett} shows the time evolution of the central density, the central temperature and the absolute value of the cooling function for the initial "polytropic" (labeled by A,B,C) and NFW (labeled by D,E,F) dark matter density profiles. In each case three different initial gas density profiles determined by the radii of the halo boundary are considered (see corresponding labels at plot A and D). 

The initial boundary radii chosen as small as ($R_H=16 pc$ and $R_H=20 pc$)
correspond to steep initial gas profiles. Due to cooling the collapse happens
within a relatively short time interval. After an initial period of strong 
cooling, the cooling function $\Lambda$ (see plots C and F) decreases due to
the decline of temperature but rises again catastrophically if the central density increases.

For shallower initial density profiles corresponding to $R_H=20 pc$ and
$R_H=28 pc$  after a first short contraction phase, an intermediate delayed
contraction phase occurs. In this case, the central density is lower when
the cooling function rises again. The further increase in density is not 
sufficient to compensate the decline of temperature in order to keep the cooling function 
further increasing. Thus, the cooling function is declining before it
rises again like above and exhibits some hump. 

In the case of even shallower profiles ($R_H=24 pc$ and $R_H=34 pc$) 
the period of delayed contraction is significantly larger. 
(note the logarithmic time scale). 
Due to the initially even lower central density the cooling function exhibits a series of oscillations until a final
unlimited raise of the cooling function occurs.

In all cases, after a very short initial cooling
period the temperature is varying only very little during the contraction phase.

\begin{figure}
   \centering
   \includegraphics[width=\columnwidth]{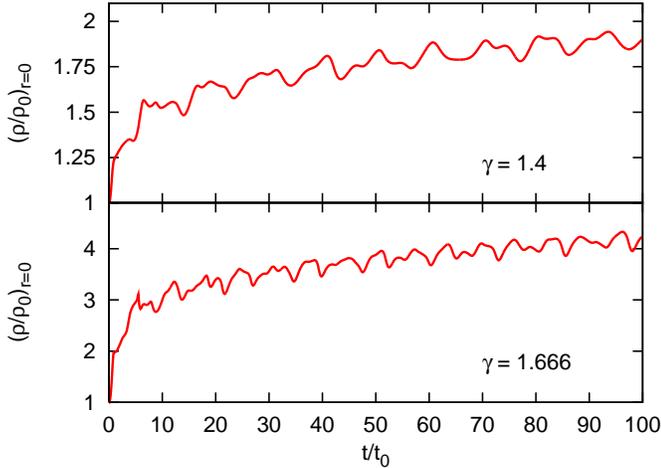}
      \caption{A blow-up of the tome evolution of the central density during the epoch of delayed collapse for two different $\gamma$}
      \label{wiggle}
   \end{figure}

The oscillations in the plots for the
shallowest profile appear as nearly periodic. Fig. \ref{wiggle} shows this in a blow-up of the plot 
during the delayed collapse for two different e.o.s. for the gas
($\gamma=[1.4, 1.666]$). The quasi-periodicity depends on the value of
$\gamma$ for the gas. 

\begin{figure}
   \centering
   \includegraphics[width=\columnwidth]{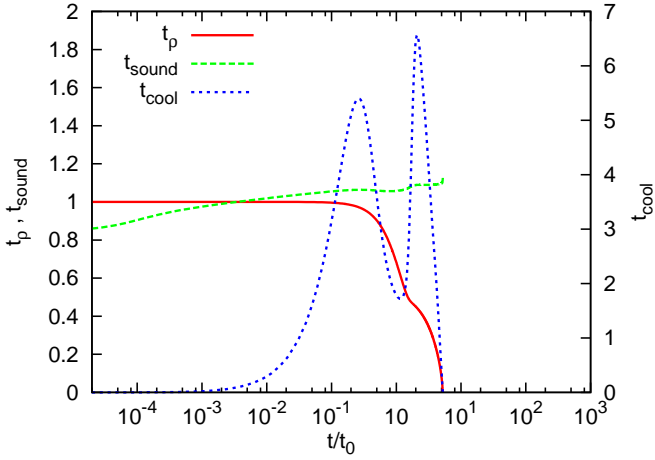}
      \caption{Time evolution of characteristic dynamic time $t_\rho$, pressure response time 
$t_{sound}\propto 1/\sqrt{T}$ and cooling time $t_{cool}$ for the initial
configuration ($R_H=20 pc$, initial "polytropic" DM configuration) 
in the central region}
      \label{times}
   \end{figure}

For an interpretation of the above described behaviour we refer to Fig
\ref{times}. This figure shows the evolution of the characteristic dynamic
time $t_\rho$, pressure response time $t_{sound}\propto 1/\sqrt{T}$ and
cooling time $t_{cool}$ for the intermediate configuration ($R_H=20 pc$,
initial "polytropic" DM configuration) in the central region. We notice that after a short cooling period the pressure is not longer able to balance the gravitational force: the dynamical time $t_\rho$ becomes smaller than all other characteristic times and the system undergoes contraction without any pressure resistance. The gas falls almost freely. If the initial density is not too high the collapse gets delayed due to again increasing pressure force (the net acceleration becomes less than for pure gravitational infall). However, at nearly constant temperature the density enhancement due to contraction enforces the cooling process again (the cooling time decreases till a minimum) in a way that pressure resistance breaks down and the gravitational collaps will be enforced. In the discussed case, this delay happens only once. Then the configuration is driven into the cooling catastrophe, i.e. both the cooling  and the dynamical times approach zero. For even shallower initial density profiles, these periods of collaps delay may happen several times.

We wanted to study the influence of any DM dynamics on the collapse behaviour
of the gas. To this end we repeated the above described  calculations under
the assumption that the distribution of the DM remains static, i.e. the DM
appears as a source of an additional gravitational potential, only. 
This approximation is often used for only studying the dynamical evolution of the gas embedded in the DM halo (see, e.g., \cite{conroy07}).

\begin{figure}
   \centering
   \includegraphics[width=\columnwidth]{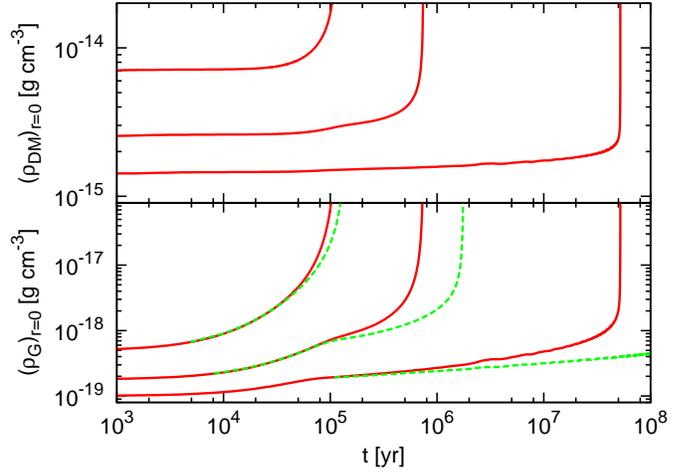}
      \caption{The figure shows the time evolution of the central gas and DM density in comparison to the case of an assumed static DM distribution (dashed line) for an initial NFW profile. At x-axis the true time is given.}
      \label{evo-nfw}
   \end{figure}

\begin{figure}
   \centering
   \includegraphics[width=\columnwidth]{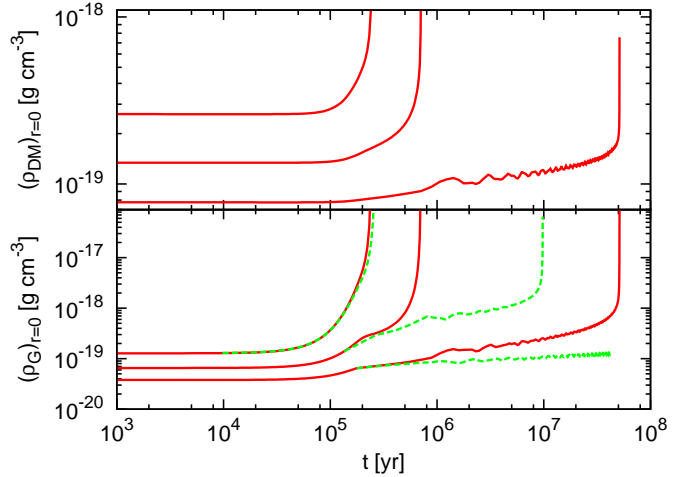}
      \caption{The same as in Fig. \ref{evo-nfw} but for an initial "polytropic" DM distribution.}
      \label{evo-poly}
   \end{figure}

In Figs. \ref{evo-nfw}  we show the time evolution of the central density in
true units for initial NFW DM configurations with $R_H=[20,28,34] pc$ (higher
initial central density corresponds to smaller initial $R_H$). The upper panel
shows the evolution of the DM central densities and the lower panel shows the
evolution of central gas densities. The dashed lines in the lower panel show
the gas evolution for the case of static DM configuration. It can be clearly
noticed that the evolution goes much faster, in general, if the dynamical
evolution of the DM is taken into account. The shallower the initial density
profile is the larger the amplification effect appears and the more extended the period of delayed collapse is. 
In Fig. \ref{evo-poly} the analogous
results for the case of an initial "polytropic" DM configuration is presented ($R_H=[20,28,34] pc$). Note the appearance of the quasi-oscillations also for the evolution of the central DM density.


\begin{figure}
   \centering
   \includegraphics[width=\columnwidth]{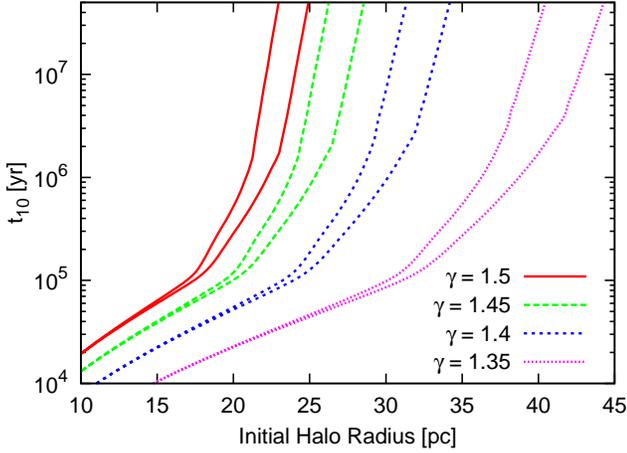}
      \caption{The collapse time $t_{10}$ is defined as the moment when the
        central gas density $\rho$ reaches $\approx 10 \rho_{ini}$.  $t_{10}$
        is shown as function of the initial halo radii (for a fixed mass $M_H
        = 10^6 M_{\odot} $). The initial density for the DM follows the
        NFW-profile. For each $\gamma$ the left-hand graph corresponds to
        a static DM distribution, while the right-hand graph corresponds to the
        dynamic DM evolution.}
      \label{c-time-nfw}
   \end{figure}

A measure for the collapse strength could be the period of time $t_{10}$ during
which the central density was increasing by a factor of 10 with respect to the
initial value $\rho_0$. The Fig. \ref{c-time-nfw} shows for different
polytropic indices the dependence of $t_{10}$ on the chosen initial radii in
the case of an initial NFW DM density profile. For given mass, the initial
radii $R_H$ are a measure for the steepness of the initial density
profiles. The line style indicates on the chosen value for $\gamma$, 
being the ratio of specific heats for the gas. Note, the effective $\gamma_{DM}$ for the DM is fixed to be $\gamma_{DM}= 5 /3$, throughout. 

It can be
noticed that increasing $R_H$ leads to an increasing collapse time. A larger
$\gamma$ for the gas, i.e. a stiffer e.o.s. of the gas, leads to a larger $t_{10}$ for given
$R_H$. The figure shows pairwise plotted graphs. In each case, the
lefthand-sided graph shows the dependence in case of a static DM
distribution. This means that taking into account a dynamical DM evolution
coupled to the dynamics of the gas enhances the contraction, i.e. leads to
shorter $t_{10}$. One can distinguish three different segments in each
  graph. Each segment corresponds to a different type of collapse behaviour described
  above. The transitions between these different collapse phases is characterized by tight intervals of values $R_H$.  

\begin{figure}
   \centering
   \includegraphics[width=\columnwidth]{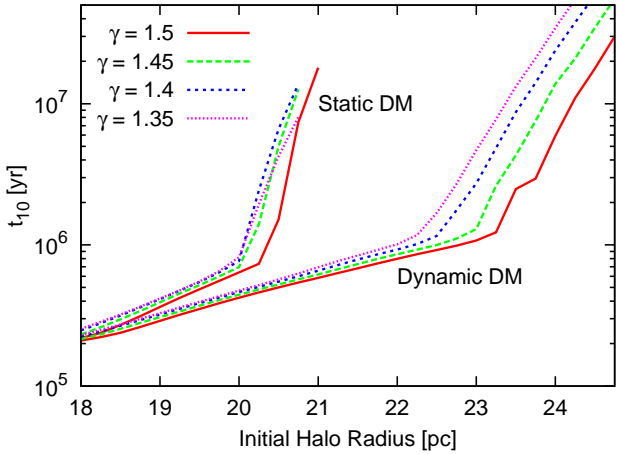}
      \caption{The same as in Fig. \ref{c-time-nfw} but for initial "polytropic" DM distributions.}
      \label{c-time-poly}
   \end{figure}

Fig. \ref{c-time-poly} shows the dependences in the case of an initial
"polytropic" DM density profile. Here, the static DM distribution leads to
much longer $t_{10}$ for given initial radii $R_H$ in comparison with the case
of dynamical interaction between gas and DM (labeled by "Dynamic DM" in the
plot). Interestingly, for the dynamic case the dependence on the polytropic
index is inverse with respect to the above case of an initial NFW
profile. Here, the three collapse phases seen above degenerate into two,
but the transition between them happens within a much narrower range of $R_H$.
In the case of the dynamic DM a break point at radii
$R_H \approx 22 - 23 pc$ separates the short-time collapse from the delayed
collapse.

.   

\section{Conclusions}

Our goal was to follow the time-dependent behaviour of a spherical halo consisting of DM
and gas with high accuracy in a 1D Model. In particular, we wanted to study the consequences of the dynamic interaction between the two kinds of matter onto the inner matter distribution if the gas is allowed to dissipate energy. 

We made the assumption that the DM is dynamically reacting fast enough on any change of the overall gravitational potential via processes of violent relaxation. This roughly justifies our crude fluid approximation for the DM which actually describes the adiabatic behaviour of a collisionless particle configuration under selfgravitation and being coupled to an embedded dissipating gas cloud via gravitational interaction only. 

Our results show that the dynamical interaction of gas and DM leads to an
enhancement of the collapse behaviour of the system (in agreement with the results of \cite{kazantzidis}). The characteristic times
for the contraction process due to the dissipation (cooling) of the gas
component are very sensitive with respect to the initial density profiles
although the configurations are all in a stable hydrostatic equilibrium
initially. The steeper the central density profiles are the shorter the
characteristic collapse times are.  For halo masses of $\approx M_6=10^6
M_{\odot}$ examined by us, two collapse phases for the gas fraction can be distinguished: After an initial short-period contraction follows a quasi-stationary state the duration of which depends on the initial density profile and possibly on the actual cooling function.

We have considered the processes of collisional ionisation and hydrogen recombination cooling, only. Therefore, the cooling function $\Lambda$ vanishes effectively below $T\approx 10^4$ K. However, for halo masses of about $M_6=10^6
M_{\odot}$ molecular hydrogen cooling may be important. Detailed simulations by \cite{yoshida03} and \cite{yoshida06} show that this leads indeed to further gas collapse and eventually to the formation of first stars in the central region. In this context, the authors stress the point that the processes as cooling and cloud formation are mainly determined by the particular dynamics of the gravitational collapse. Thus, including molecular hydrogen cooling will lead to an even steeper gas density profile and hence is expected to strengthen the collapse behaviour of the whole considered configuration.

Our result may also enforce the problem described by \cite{conroy07}. Their outcome is that only a sensitive fine tuning of different cooling and heating processes is able to stabilize configurations like clusters over a time comparable with the Hubble time, at least. Taking into account the dynamical interaction between thermo-dynamically evolving gas and the DM halo such a fine tuning may be difficult to adjust. Especially within the central region of the considered configurations the dynamic interaction between gas and DM may alter the situation considerably.

Our computations do not indicate on any situation that the innermost DM distribution might get shallower due to interaction with the gas. Thus, including self-consistently the gravitational interaction between DM and gas is not yet sufficient to solve the discrepancy
between the NFW density profile for DM halos in simulations and the indications on a core distribution from observed rotational curves of galaxies. The tendency of even steepening the inner DM profile by gas cooling may have influence on the investigations of DM annihilation signals.

Whenever the cooling time approaches the dynamical time and both are becoming
fast decreasing the cooling catastrophe cannot be avoided. Although we have
considered only the example of cooling by recombination, this seems the
outcome for all cooling functions $\propto \rho^2$ (s. also \cite{conroy07}),
i.e. if the density is increasing the cooling time always decreases. 

\begin{figure}
   \centering
   \includegraphics[width=\columnwidth]{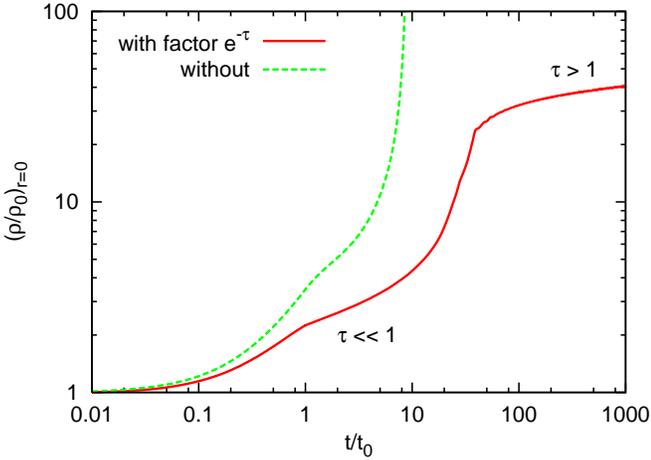}
      \caption{An example for the modified collapse behaviour if the cooling time strongly decreases with increasing optical
   depth. For comparison, the curve on the left hand shows the central density evolution for a optically thin medium.}
      \label{damping}
   \end{figure}

These considerations assume that any dissipated energy can leave the system
immediately. If the medium gets sufficiently opaque with respect to
interaction of the radiation with matter (e.g. photon-electron
scattering) the radiation may need considerable time for carrying away the
dissipated energy which may increase the cooling time considerably. To
illustrate that we have introduced a factor $\propto \exp(-n \, \sigma_T \,
\lambda)$ in front of the cooling function. Here $\lambda$ denotes some
characteristic length, $\sigma$ the relevant cross section and $n$ the number
density, e.g., of electrons. At large enough densities $n$ the optical depth
$\tau = -n \sigma_T \lambda$ may approach unity or may become even larger. In
that case the cooling time is not longer decreasing at increasing density and
one could expect that the epoch of delayed collapse may end at a stationary
configuration for particular cases. An example is given in Fig.\ref{damping}
in comparison with the above considered optically thin case. This may indicate
on a possible self-regulation process for halos within a certain mass range.
Above all, this concerns dense objects as stars but possibly also the behaviour of the very central regions of forming cosmic structures.

Our results demonstrate that the gravitational interaction between gas and DM
may have important influence on the dynamical evolution of the halo systems
during cosmic evolution.
 We have not included the possibility of continuous
matter infall/accretion onto the evolving gas-DM halos. Furtheron, we did not include non-isotropic velocity dispersions and a non-vanishing angular momentum. Considering this may alter the situation (see, e.g., recent work by \cite{tonini}) and will be subject to forthcoming research.  

\begin{appendix}

\section{Numerical Algorithms}\label{algo}

In the first step, velocity $u^*$ and pressure $p^*$ at the cell interfaces
are determined by an exact Riemann solver. With these values the position 
of the cell interfaces after half of the timestep are computed:
\begin{equation}
r^\prime_{i+\frac{1}{2}} =
r_{i+\frac{1}{2}} \left( t\right) + 
\frac{1}{2} \, \Delta t \; u^*_{i+\frac{1}{2}}
\end{equation}
Because the mass coordinates of DM and gas are independent, the 
gravitational acceleration for the DM is computed by interpolating
the enclosed mass of the gas using the trapezoid rule:
\begin{eqnarray}
\Big[ g_{DM} \Big]_i &=& 
4 \, \pi \, G \;
\frac{ 
  \left[m_{DM}\right]_i +
  \left[m_{G}\right]_i
}
{
  \bar{r}^2_i}
\end{eqnarray}
with:
\begin{eqnarray}
\bar{r}_i &=&
 \frac{ 
    [ r^\prime_{DM} ]_{i+\frac{1}{2}} +
    [ r^\prime_{DM} ]_{i-\frac{1}{2}}
}{2}
\nonumber\\
\left[m_{DM}\right]_i &=& 
\frac{ 
  \left[m_{DM}\right]_{i+\frac{1}{2}} + \left[m_{DM}\right]_{i-\frac{1}{2}}
}{2}
\nonumber\\
\left[m_{G}\right]_i &=& 
\left[m_{G}\right]_{j+\frac{1}{2}} + 
\left( \bar{r} - [r^\prime_G]_{j+\frac{1}{2}} \right)
\frac{ 
  [m_{G}]_{j+\frac{3}{2}} -
  [m_{G}]_{j+\frac{1}{2}}}
{
  [r^\prime_{G}]_{j+\frac{3}{2}} -
  [r^\prime_{G}]_{j+\frac{1}{2}}}
\end{eqnarray}
with the gas cell $j$, being the outermost gas cell enclosed 
($[\bar{r}_{DM}]_i > \, [r^\prime_G]_{j+\frac{1}{2}} 
> \, [\bar{r}_{DM}]_{i-1})$. The procedure to calculate the 
gravitational acceleration for the gas is analogous.
For computation of the cooling function the temperature is computed 
according to
\begin{equation}
T_i = \frac{m_H}{k_B} \, p_i \, \tau_i
\end{equation}
Then, the cooling function $[\Lambda \, \tau]_i$ can be calculated using
Equations (\ref{black}), (\ref{cool3}) and (\ref{realcool}). To conclude the
timestep, the quantities at the time $t+\Delta t$ are now computed (in this order):
\begin{eqnarray}
r_{i+\frac{1}{2}} \left( t+\Delta t \right) &=& 
r_{i+\frac{1}{2}} \left( t\right) + u^*_{i+\frac{1}{2}} \cdot \Delta t
\nonumber \\
\tau_i \left( t+\Delta t \right) &=& \frac{r^3_{i+\frac{1}{2}}}{3} - 
\frac{r^3_{i-\frac{1}{2}}}{3}
\nonumber \\
u_i \left( t+\Delta t \right)
&=& u_i \left( t \right) - \, \frac{\Delta t}{\Delta m_i} \,
\Bigg[  
  \left[r^{\prime 2} p^* \right]_{i+\frac{1}{2}} -\,
  \left[r^{\prime 2} p^* \right]_{i-\frac{1}{2}}
\nonumber \\
&& 
  - \,\frac{p^*_{i+\frac{1}{2}} + p^*_{i-\frac{1}{2}}}{2}
  \left( r^\prime 2_{i+\frac{1}{2}} - r^\prime 2_{i-\frac{1}{2}} \right)
\Bigg] 
+ \, \Delta t \; g_i
\nonumber \\
\varepsilon_i \left( t+ \Delta t \right)
&=& \varepsilon_i \left( t \right) - \, \frac{\Delta t}{\Delta m_i} \,
\Bigg[
\left[r^{\prime 2} p^* u^* \right]_{i+\frac{1}{2}} - \,
\left[r^{\prime 2} p^* u^* \right]_{i-\frac{1}{2}}
\Bigg]
\nonumber \\
&& + \, \Delta t \; \Bigg[\frac{u^*_{i+\frac{1}{2}} + u^*_{i-\frac{1}{2}}}{2}
\, g_i + \left[\Lambda \, \tau \right]_i \Bigg]
\end{eqnarray}
The boundary condition at $r = 0$ is realized by a reflecting boundary. 
Central pressure and velocity
are computed using the solution of the Riemann-problem between the innermost 
cell and a ghost cell with the same density and pressure, 
but with a negative velocity. The boundary between gas and vacuum is treated by
solving the vacuum Riemann problem (see \cite{toro}, p. 138). 
The solution consists only of one rarefaction wave ending at the contact
discontinuity. Thus, the pressure is $p^\star = 0$., there. 
Using the Lagrangian
version of the Generalized Riemann Invariant (see \cite{muscl}, p. 105):
\begin{equation}
(u^\star - u) \mp \frac{2}{\gamma - 1} 
\left(\tau^\star a^\star - \tau \, a \right) = 0
\end{equation}
one gets for the velocity:
\begin{equation}
u^\star = u_{i_{max}} 
+ \frac{2}{\gamma - 1} \, \tau_{i_{max}} \,  a_{i_{max}}
\end{equation}
with the Lagrangian speed of sound $a = \sqrt{\gamma  p / \tau}$.

Instead of a sharp cut-off, the NFW-profile reaches zero only for 
$r\rightarrow\infty$. We use a reflecting boundary condition there to inhibit
any artificial inflow or outflow. Since the outer boundary of the dark-matter 
profile is much more extended than for the gas, this choice should not affect the
dynamics of the central region.
\end{appendix}

\begin{acknowledgements}
We thank Naoki Yoshida for giving us many constructive comments and helpful
suggestion. J.K. was supported by the Deutsche Forschungsgemeindschaft
under the project MU 1020/6-4.s. 
\end{acknowledgements}


\begin{thebibliography}{}

\bibitem[Ascasibar et al. (2006)]{acasibar06} 
  Ascasibar,Y., Jean, P., Boehm, C., Kn\"odlseder, J., 2006, MNRAS, 368, 1695

\bibitem[Binney and Tremaine (1988)]{bt} 
  Binney, J. and Tremaine, S., Galactic Dynamics,
  Princeton University Press, 1987

\bibitem[Blumenthal et al. (1986)]{blumenthal86} 
  Blumenthal, G.R., Faber, S.M., Flores, R., Primack, J.R., 
  1986, ApJ, 301, 27-34
 
\bibitem[Black (1981)]{black} Black, J.H., 1981, MNRAS, 197, 553-563

\bibitem[Cardone \& Sereno (2005)]{cardone05} 
  Cardone, V.F., Sereno, M., 2005, Astron.Astrophys. 438, 545

\bibitem[Chi\`eze et al. (1997)]{chieze} 
  Chi\`eze, J.-P., Teyssier, R., Alimi, J.-M., 1997, ApJ, 484, 40

\bibitem[Conroy \& Ostriker (2007)]{conroy07} 
  Conroy, C., Ostriker, J.P., 2007, astro-ph/07120824

\bibitem[Diemand et al. (2007)]{diemand07} 
  Diemand, J., Kuhlen, M., Madau, P., 2007, ApJ, 657, 262

\bibitem[Dutton et al. (2007)]{dutton07} 
  Dutton, A.A., van den Bosch, F.C., Dekel, A., Courteau, S.Cardone, V.F., 
  2007, ApJ, 654, 27

\bibitem[Eggen, Lynden-Bell \& Sandage (1962)]{eggen62} 
  Eggen, O.J., Lynden-Bell, D., Sandage, A.R., 1962, ApJ, 136, 748

\bibitem[Gentile et al. (2004)]{gentile04} Gentile, G., Salucci, P., Klein, U., Vergani, D., Kalberla, P., 2004, MNRAS 351,903

\bibitem[Gnedin et al. (2004)]{gnedin04}
  Gnedin, O.Y., Kravtsov, A.V., Klypin, A.A., Nagai, D., 2004, ApJ, 616, 16

\bibitem[Hoeft, M{\"u}cket\&Gottl{\"o}ber (2004)]{hoeft04} 
  Hoeft, M., M\"ucket, J.P., Gottl\"ober, S., 2004, ApJ, 602, 162

\bibitem[Jesseit, Naab \& Burkert (2002)]{jesseit02}
  Jesseit, R., Naab, T., Burkert, A., 2002, ApJ, 571, L89

\bibitem[Kazantzidis et al. (2006)]{kazantzidis}
  Kazantzidis, S., Kravtsov, A.V., Zentner A.R., Allgood, B., Nagai,
  D., Moore, B., 2004, ApJ, 611, L73

\bibitem[Kleinheinrich et al. (2006)]{kleinheinrich06} 
  Kleinheinrich, M., Schneider, P., Rix, H.-W., Erben, T., Wolf, C., Schirmer,
  M., Meisenheimer, K., Borch, A., Dye, S., Kovacs, Z., Wisotzki, L, 
  2006, Astron.Astrophys. 455, 441

\bibitem[Kull et al. (1997)]{kull} 
  Kull, A., Treumann, R.A., B\"oringer, H., 1997, ApJ, 484, 58-62

\bibitem[Kuzio et al. (2006)]{kuzio06} 
  Kuzio de Naray1, R., McGaugh1, S.S., de Blok1, W. J. G., Bosma, A., 2006, 
  ApJS, 165, 461

\bibitem[van Leer (1979)]{muscl} 
  van Leer, B., 1997, Jounal of Computational Physics, 32, 101-136

\bibitem[Lyndon-Bell (1967)]{lb} 
  Lyndon-Bell, D. 1967, MNRAS, 136, 101

\bibitem[Marchesini et al. (2002)]{marchesini02}  
  Marchesini, D., D'Onghia, E., Chincarini, G., Firmani, C., Conconi, P., 
  Molinari, E., Zacchei, A., 
  2002, ApJ, 575, 801

\bibitem[Navarro, Frenk \& White (1997)]{navarro97} 
  Navarro, J.F., Frenk, C.S., White, S.D.M., 1997, ApJ, 490, 493

\bibitem[Persic et al. (1996)]{persic96} Persic, M., Salucci, P., Stel, F., 1996, MNRAS, 281, 27

\bibitem[Ryden \& Gunn (1987)]{ryden87} 
  Ryden, B.S., Gunn, J.E., 1987, ApJ, 318, 15

\bibitem[Salucci et al. (2007)]{salucci07} Salucci, P., Lapi, A., Tonini, C., Gentile, G., Yegorova, I., Klein, 
U., 2007, MNRAS, 378, 41

\bibitem[Salucci et al. (2003)]{salucci03} Salucci, P., Walter, F., Borriello, A., 2003, Astron.Astrophys., 409, 53

\bibitem[Sellwood \& McGaugh (2005)]{sellwood05} 
  Sellwood, J.A., McGaugh, S.S., 2005, ApJ, 634, 70

\bibitem[Stoehr et al. (2003)]{stoehr03} 
  Stoehr, F., White, S.D.M., Springel, V., Tormen, G., Yoshida, N., 2003, 
  MNRAS, 345, 1313

\bibitem[Taylor \& Navarro (2001)]{taylor01} 
  Taylor, J.E., Navarro, J.F., 2001, ApJ, 563, 483

\bibitem[Teyssier et al. (1997)]{teyssier} 
  Teyssier, R., Chi\`eze, J.-P.,  Alimi, J.-M., 1997, ApJ, 484, 36

\bibitem[Toro (1999)]{toro} 
  Toro, E., Riemann Solvers and Numerical Methods for Fluid Dynamics 
  (2nd Edition),  Springer-Verlag Berlin Heidelberg, 1999

\bibitem[Tonini et al. (2006)]{tonini}
  Tonini, C., Lapi, A., Salucci, P., 2006, 649, 591

\bibitem[Vasiliev (2006)]{vasiliev06} 
  Vasiliev E., 2006, JETO Letters 84/2, astro-ph/0601669

\bibitem[Yoshida et al. (2003)]{yoshida03} 
  Yoshida, N., Abel, T., Hernquist, L., Sugiyama, N., 2003, ApJ, 592, 645

\bibitem[Yoshida et al. (2006)]{yoshida06} 
  Yoshida, N., Kazuyuki, O., Hernquist, L., Abel, T., 2006, ApJ, 652, 6

\bibitem[Young (1980)]{young80} Young P., 1980, ApJ, 242, 1232

\bibitem[Zeldovich et al. (1980)]{zeldovich80} 
  Zeldovich, Y.B., Klypin, A.A., Khlopov, M.Y., Chechetkin, V.M. 1980,
  Soviet J.Nucl.Phys., 31, 664

\end{thebibliography}
\end{document}